\documentclass[preprint,preprintnumbers,amsmath,amssymb,superscriptaddress]{revtex4}

\usepackage{graphicx}
\usepackage{dcolumn}
\usepackage{bm}

\begin{document}

\preprint{APS/123-QED}

\title{Systematic control of surface Dirac fermion density on topological insulator Bi$_2$Te$_3$}

\author{Y. Xia}
\affiliation{Joseph Henry Laboratories of Physics, Department of Physics, Princeton University, Princeton, NJ 08544, USA}
\author{D. Qian}
\affiliation{Joseph Henry Laboratories of Physics, Department of Physics, Princeton University, Princeton, NJ 08544, USA}
\affiliation{Department of Physics, Shanghai Jiao Tong University,
Shanghai 200030, China}
\author{D. Hsieh} \affiliation{Joseph Henry
Laboratories of Physics,  Department of Physics, Princeton
University, Princeton, NJ 08544, USA}
\author{R. Shankar}
\affiliation{Joseph Henry Laboratories of Physics, Department of Physics, Princeton University, Princeton, NJ 08544, USA}
\author{H. Lin}
\affiliation{Department of Physics, Northeastern University, Boston,
MA}
\author{A. Bansil}
\affiliation{Department of Physics, Northeastern University, Boston,
MA}
\author{A.V. Fedorov}
\affiliation{Advanced Light Source, Lawrence Berkeley National Laboratory, University of California, Berkeley, CA 94305, USA}
\author{D. Grauer}
\affiliation{Department of Chemistry, Princeton University, Princeton, NJ 08544, USA}
\author{Y. S. Hor}
\affiliation{Department of Chemistry, Princeton University, Princeton, NJ 08544, USA}
\author{R. J. Cava}
\affiliation{Department of Chemistry, Princeton University, Princeton, NJ 08544, USA}
\author{M. Z. Hasan}
\affiliation{Joseph Henry Laboratories of Physics,  Department of Physics, Princeton University, Princeton, NJ 08544, USA}
\affiliation{Princeton Center for Complex Materials, Princeton University, Princeton, NJ 08544, USA}
\affiliation{Princeton Institute for the Science and Technology of Materials, Princeton University, Princeton, NJ 08544, USA}

\date{\today}


\begin{abstract}

Three dimensional (3D) topological insulators are quantum materials with a spin-orbit induced bulk insulating gap that exhibit quantum-Hall-like phenomena in the absence of applied  magnetic fields. They feature surface states that are topologically protected against scattering by time reversal symmetry. The proposed applications of topological insulators in device geometries rely on the ability to tune the chemical potential on their surfaces in the vicinity of the Dirac node. Here, we demonstrate a suite of surface control methods based on a combination of photo-doping and molecular-doping to systematically tune the Dirac fermion density on the topological (111) surface of \textbf{Bi$_2$Te$_3$}. Their efficacy is demonstrated via direct electronic structure topology measurements using high resolution angle-resolved photoemission spectroscopy. These results open up new opportunities for probing topological behavior of Dirac electrons in \textbf{Bi$_2$Te$_3$}. At least one of the methods demonstrated here can be successfully applied to other topological insulators such as the bulk-insulating-Bi$_{1-x}$Sb$_x$, Sb$_2$Te$_3$ and Bi$_2$Se$_3$ which will be shown elsewhere. More importantly, our methods of topological surface state manipulation demonstrated here are highly suitable for future spectroscopic studies of topological phenomena which will complement the transport results gained from the traditional electrical gating techniques.

\end{abstract}

\maketitle

Recently, there has been a surge of research interest in the newly discovered
class of 3D time reversal invariant topological insulators
\cite{MooreNatphys, zhang, kane, Fuprl, Mooreprb, Fuprb} both from the fundamental physics and potential application viewpoints. Such systems have been proposed to exhibit many exotic quantum phenomena \cite{Qi, Fuarxiv,Seradjeh} that can be realized in condensed-matter table-top settings. Three dimensional topological insulator phases with gapless topological edge states have been observed in Bi$_{1-x}$Sb$_x$ \cite{Hsiehnature, Hsiehscience}, Bi$_2$Se$_3$ \cite{Xia} and Bi$_2$Te$_3$ \cite{noh, Hsieharxiv, chen}. Topological order (the physically observable quantity that uniquely characterizes the new state, and in some sense comes closest to an "order parameter" of the state in the absence of the spontaneously broken symmetry unlike that in a superconductor) in a topological insulator cannot be determined without spin-texture or Berry's phase imaging measurements. Using spin-resolved methods, Hsieh \emph{et.al.} measured the Berry's phase texture of the Bi$_{1-x}$Sb$_x$ and Bi$_2$Te$_3$ classes of material \cite{Hsiehscience, Hsieharxiv}, confirming the Fu-Kane-Mele type Z$_2$ topological order \cite{Fuprl, Mooreprb, Fuprb} in these materials. In the momentum space, the pairs of time-reversal invariant edge states can be described as massless Dirac fermions inside the bulk band-gap. A strong topological insulator (STI) is the one characterized by having an odd number pair of such states, which are topologically protected against backscattering. The STI phase is a topologically distinct novel phase of matter and cannot be adiabatically reduced to the quantum spin Hall phase proposed in the strongly spin-orbit coupled two dimensional electron gases \cite{bernevig}. It is the surface dynamics and control of topological Dirac fermions in the Bi$_2$Te$_3$ class that we focus in this paper.

While the density of state at the Fermi level (DOS(E$_F$))in a conventional quantum Hall system (QHE in 2DEG) can be tuned via an external magnetic field, this method cannot be applied to the STIs for a magnetic field breaks time reversal symmetry and the topological protection is thus lost. Furthermore, chemical techniques such as the bulk doping to reach the Dirac point are undesirable as they introduce extra carriers which either turn the bulk into a semi-metal or an alloy or increase the residual conductivity beyond the surface conduction limit. None of these conditions are desirable for a functional topological insulator. Here, we demonstrate a suite of surface doping methods which can tune the carrier density on the surfaces of the topological insulator system Bi$_2$Te$_3$ in a time-reversal invariant way. Using angle-resolved photoemission spectroscopy, we demonstrate the capability of electronic structure engineering of the Bi$_2$Te$_3$ surface with alkali atom (potassium, K) deposition, molecular adsorption and photo-doping without breaking the time reversal or the Z$_2$ invariance. We show that the chemical potential can be systematically tuned to shift the surface Dirac point downwards in energy. Moreover, we report that the effect of NO$_2$ molecular adsorption can be further manipulated by \textit{photon-assisted stimulation}, thereby controlling the surface carrier density in the opposite directions. These methods of tuning carrier densities on the surface of a topological insulator open up new research avenues for spintronic and quantum computing applications where the control of the density of Dirac fermions on the surfaces is critical to observe many spin-orbit or spintronic quantum effects. More importantly, our methods of topological surface state manipulation, highly suitable for spectroscopic studies, demonstrated here will complement the traditional electrical gating techniques.

Single crystal of Bi$_2$Te$_3$ was grown by melting stoichiometric mixtures of high purity elemental Bi (99.999\%) and Te (99.999\%) in a vacuum sealed 4 mm inner diameter quartz tube at 800 $^{\circ}$C. The sample was then cooled over a period of two days to 550 $^{\circ}$C, and then annealed at that temperature for 5 days. A single crystal with lowest residual conductivity characterized via transport methods was obtained. Single crystals could be easily cleaved along the basal plane. The crystal was confirmed to be
single phase and identified as having the rhombohedral crystal structure by X-ray powder diffractometry using a Bruker D8 diffractometer with Cu K$\alpha$ radiation and a graphite diffracted beam monochromator.

High-resolution ARPES measurements were then performed using 30 to 55eV photons at Beamline 12.0.1 of the Advanced Light Source in Lawrence Berkeley National Laboratory. The energy and momentum resolution were typically 15meV and 1.5\% of the surface Brillouin Zone (BZ) respectively using a Scienta analyzer. The in-plane crystal orientation was determined by Laue x-ray diffraction prior to inserting into the ultra-high vacuum
measurement chamber. The samples were cleaved \emph{in situ} at 10K under pressures of less than $2\times 10^{-11}$ torr, resulting in shiny flat surfaces. NO$_2$ molecular doping of the Bi$_2$Te$_3$ surface was achieved via controlled exposures to NO$_2$ gas
(Matheson, 99.5\%), after the surface was allowed to relax for 1 hour after cleavage \cite{Hsieharxiv}. The adsorption effects were studied under static flow mode: exposing the sample to the gas for a certain time then taking data after the chamber was pumped down to the base pressure. The sample temperature was maintained at 10K during exposure to NO$_2$. For K deposition, a heated K evaporation source was used operating at 5.5A. We also present band calculations for the Bi$_2$Te$_3$(111) surface with parameters optimized based on experimental data, which were performed with the LAPW method in slab geometry using the WIEN2K package \cite{blaha01}. The calculated band-structure was found to be qualitatively similar to previous report calculated using different methods\cite{hzhang}. GGA of Perdew, Burke, and Ernzerhof \cite{perdew96} was used to describe the exchange-correlation potential with SOC included as a second variational step. The surface was simulated by placing a slab of six quintuple layers in vacuum. A grid of $35\times 35 \times 1$ points were used in the calculations, equivalent to 120 k-points in the irreducible BZ and 2450 k-points in the first BZ.

\begin{figure}
\includegraphics[width=.70\textwidth]{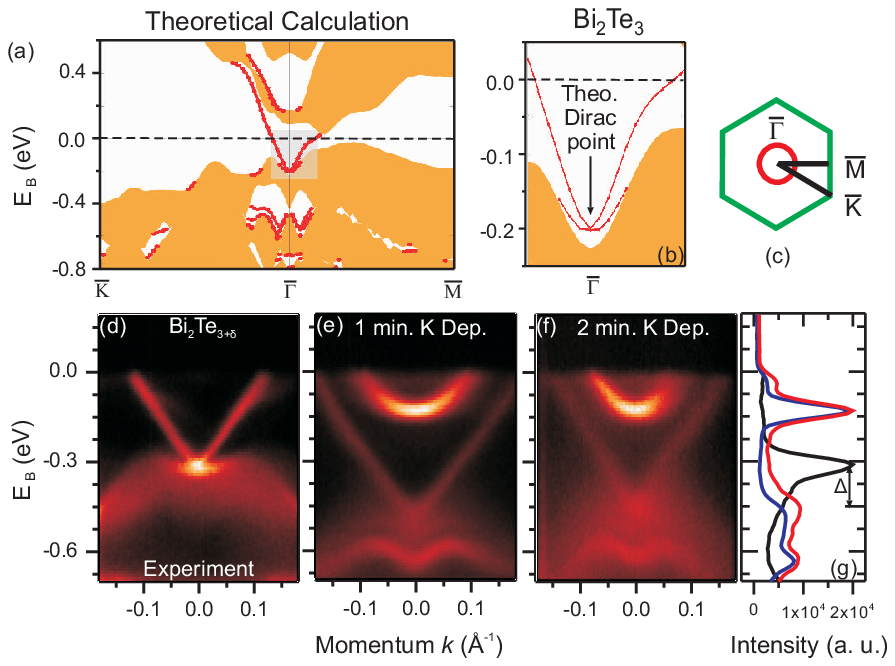}
\caption{\label{fig:fig1} \textbf{Electron doping via potassium (K) deposition on Bi$_2$Te$_3$[111] topological surface :} (a)-(b) (111) Surface band (red lines) calculation with spin-orbit coupling shows a pair of non-degenerate Dirac bands
crossing the E$_F$. The bulk band projection is denoted by the shaded
region in orange. The resulting (c) Fermi surface is a single non-degenerate ring
centered at the zone center. (d) ARPES data on Bi$_{2}$Te$_{3+\delta}$ shows a slight rise of the chemical potential relative to the stoichiometric compound Bi$_{2}$Te$_{3}$ \cite{Hsieharxiv}. The surface band structure after depositing potassium (K) for (e) one minute and (f) two minutes show additional electron-doping of the surface
states. The (g) energy distribution curves at $\bar{\Gamma}$ show
that relative to the pristine sample (black), the chemical potential
is shifted by 140meV after one minute of (red) of K deposition. Only a small additional shift of the E$_F$ position is observed upon an additional minute of K deposition (blue).}
\end{figure}

Calculated band structure along the $\bar{K}$-$\bar{\Gamma}$-$\bar{M}$ direction shows that spin-orbit coupling induces a single metallic surface band near the zone center
(Fig.~\ref{fig:fig1}(a)-(b)). The resulting surface Fermi surface is
a single non-degenerate ring centered at $\bar{\Gamma}$. Although the experimental results are in qualitative agreement with calculations and the chemical potential lies in the gap only intersecting the surface states (Fig.1), the observed differences could result from the fact that all Bi$_2$Te$_3$ samples have a relaxed cleaved surface \cite{noh, Hsieharxiv}.
Because of details of the surface relaxation (thus band bending) of Bi$_2$Te$_3$ has no simple correlation with the bulk carrier concentration (insulating or not) of our samples it is not possible to find a unique fingerprint of the bulk insulating state by just studying the surface Fermi surface shape. We have experimentally determined that in the presence of relaxation, slightly excess Te atoms added in the bulk can electron dope the surface (Figure 1(d)). These effects taken together may account for some of the differences between calculated band structure and the experimental data.

In Figure ~\ref{fig:fig1}(e)) we present ARPES data taken at 30eV after potassium (K) is deposited onto the sample for approximately one minute. The material becomes strongly electron doped, with the bottom of the "V" (Dirac band) band rigidly shifted by approximately 140meV. The Fermi velocity of the Dirac band is unchanged with doping. In addition, we observe an inner resonant-state (due to surface band bending) feature that appears to be the bottom of a pair of spin-split parabolic bands shifted from each other in k-space. These bands are reminiscent of those observed in the SS of Au(111) and Sb(111) at $\bar{\Gamma}$. Assuming that each K atom donates one electron, Luttinger count theorem suggests that the change of the electron pocket size corresponds to 7\% of a K monolayer deposited onto the sample surface. Additional K doping (Fig.~\ref{fig:fig1}(f)) has little effect on moving the chemical potential. The signal from the Dirac point however attenuates due to the increasing escape depth of the photoelectrons (Fig.~\ref{fig:fig1}(g)). Nevertheless, no gap is observed as a result of deposition, confirming that time-reversal symmetry is preserved suggesting the non-magnetic character of the valence state of K sticking onto the surface.

\begin{figure*}

\includegraphics[width=0.7\textwidth]{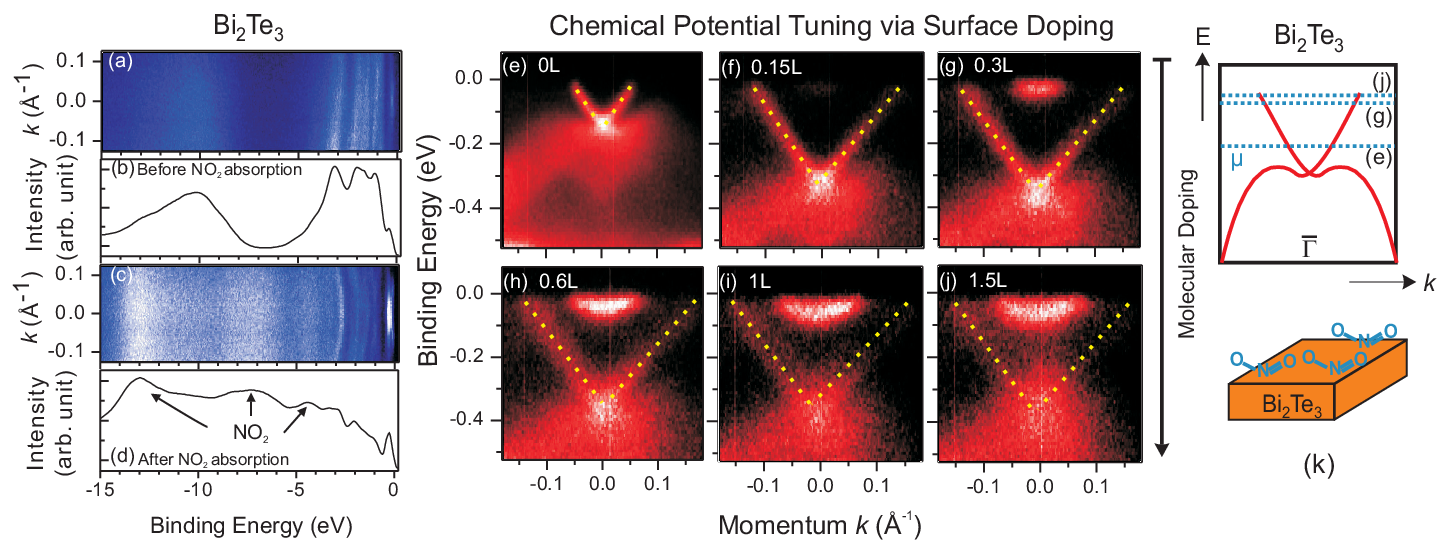}
\caption{\label{fig:fig2} \textbf{Manipulation of Bi$_2$Te$_3$[111] topological surface via molecular adsorption:} (a) Angle-resolved valence band spectra (a)
prior to and (c) post exposure to 1.5 Langmuirs of NO$_2$. Panels (b)
and (d) show the angle integrated spectra for (a) and (c),
respectively. Prior to the NO$_2$ adsorption, the valence band features are sharp
and dispersive with momentum. After exposure to NO$_2$ gas, three non-dispersive
features appear at binding energies of 4.4eV, 7.5eV and 13eV. High
resolution surface band dispersion data near the $\bar{\Gamma}$ point
suggest (e) two non-degenerate surface bands intersecting at 130meV
below E$_F$, forming a Dirac point. (f)-(j), with NO$_2$ adsorption,
the chemical potential is systematically raised, thereby introducing additional
electrons into the Bi$_2$Te$_3$ surface bands. The Dirac point moves
away from E$_F$ with increasing amounts of NO$_2$ exposure. At 1
Langmuir dosage value, the lowest surface conduction band gains electron occupation, and the Dirac point reaches 350meV below E$_F$.}
\end{figure*}

We then study the effect of NO$_2$ doping on the Bi$_2$Te$_3$ (111) surface states which has been shown to be effective in graphene \cite{8,9}. To check the adsorption or sticking behavior of the molecular dopants, figure~\ref{fig:fig2}(a)-(d) we collect valence data
with a wide binding energy range above 15eV prior and post exposure of the NO$_2$ gas. On the clean undoped samples (Fig.~\ref{fig:fig2}(a)-(b)), dispersive Bi$_2$Te$_3$ valence bands are clearly observed at binding energies from 0 to 5eV, with some additional weak features at binding energies of 9-11eV characterizing the Bi$_2$Te$_3$ matrix. Upon exposure to NO$_2$, three additional intense features appear at binding energies of 4.4eV, 7.5 eV and 13eV (Fig.~\ref{fig:fig2}(c)-(d)). These new features are non-dispersive
and independent of the choice of incident photon energy, which suggest that they are due to the adsorbed distributed molecules on the sample surface. Core level signals at similar binding energies are known to appear upon depositing NO$_2$ on carbon-based materials \cite{9}. These results confirm the sticking or adsorption of NO$_2$ on Bi$_2$Te$_3$ (111) surface.

The surface band dispersions near $\bar{\Gamma}$ with varying NO$_2$ dosages are shown in Fig.~\ref{fig:fig2}(e)-(j). After equal intervals of data collection time, one finds
two significant changes in the photoemission spectra. Firstly, due to an increasing surface roughness, the measured spectra after NO$_2$ exposure are not as sharp as that from a clean surface. The total emission intensity is reduced, as the photoelectron escape depth is increased. Secondly, after NO$_2$ adsorption, the electron pocket around $\bar{\Gamma}$ grows significantly larger as the chemical potential gradually shifts away from the surface Dirac point. After a NO$_2$ exposure of 0.15 Langmuir (0.15L)
(Fig.~\ref{fig:fig2}(f)), the chemical potential is raised by approximately 170meV. The surface conduction bands gain partial occupation, introducing a weak resonant state feature near E$_F$. The pure surface band velocity is unchanged after NO$_2$ exposure. However, the surface Dirac point (degenerate Kramer's point) is rigidly shifted downwards to 300meV below E$_F$. Increasing amounts of NO$_2$ doping increase the electron carrier density on the Bi$_2$Te$_3$ surface. Eventually, after an exposure of 1L of
NO$_2$(Fig.~\ref{fig:fig2}(i)), the surface Dirac point reaches 350meV below E$_F$. Inside the non-degenerate Dirac cone, a pair of spin-split parabolic bands from the previously unoccupied surface conduction band is also observed at around 60meV. Additional exposures deteriorates the sample significantly. While the position of the Dirac point becomes
difficult to locate, judging from the size of the lowest energy electron pocket the shift of the Fermi level is minimal with further dosages. This observation is in direct contrast with that observed in graphene \cite{9}, where electrons are transferred from
the graphene layer to the NO$_2$ molecules.

We then estimate the amount of charge carrier transfer as a result of NO$_2$ deposition by applying the Luttinger count theorem. Relaxing the sample for 1 hour after cleavage, the charge concentration (figure~\ref{fig:fig2}(e)) is approximately 0.00335 electrons per unit cell. After exposure to 1L of NO$_2$ gas (Fig.~\ref{fig:fig2}(i)), the surface becomes strongly electron doped and approximately 0.039 electrons are transferred to the
surface. The mechanism behind this charge transfer is presently unclear which makes it difficult to precisely estimate the amount of NO$_2$ molecules that actually dopes the surface. Nevertheless, it is clear that a systematic tuning of the chemical potential is possible with K doping.

\begin{figure*}
\includegraphics[width=1.0\textwidth]{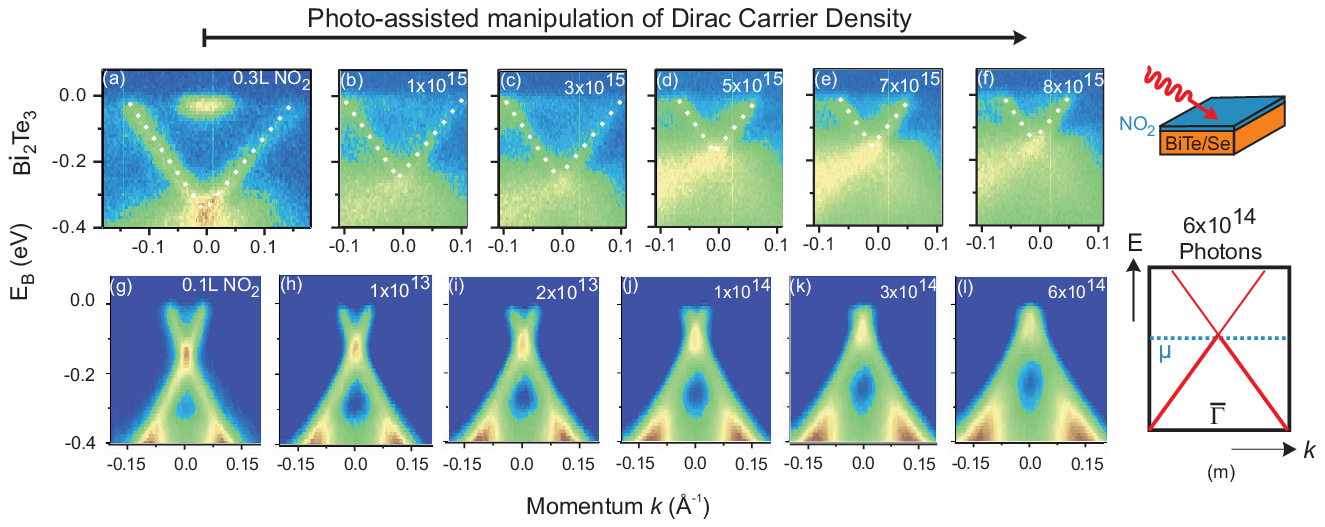}
\caption{\label{fig:fig3} \textbf{Photo-induced manipulation of the topological surface
states on Bi$_2$Te$_3$:} (a) Upon exposures of 0.3L of NO$_2$ on Bi$_2$Te$_3$, the
Dirac point is located at approximately 330meV below E$_F$. (b)-(f) Photon
exposure reverses the electron transfer to the Bi$_2$Te$_3$ surface,
moving the surface chemical potential in a well-controlled manner
shifting towards the Dirac point. With a dosage total of (b) 1$\times$$10^{15}$ photons, the Dirac point is moved to 250meV below E$_F$. The Dirac point
eventually reaches the energy-location prior to the NO$_2$ exposure with a dosage
of 8$\times$$10^{15}$ photons, which is at about 130meV below E$_F$. For comparison, the
photo-induced doping of Bi$_2$Se$_3$ surface states exposed with (g) 0.1L of
NO$_2$ gas is also presented in (h)-(l). With a dosage level of
(i)2$\times$$10^{13}$ photons, the Dirac point is shifted by 50meV to
to 100meV below E$_F$. The chemical potential is eventually stabilized at the Dirac point by the application of $6\times$$10^{14}$ photons per mm$^2$.}
\end{figure*}

We have further investigated the effect of photo-induced changes on the surface of
Bi$_2$Te$_3$. In our experiment, we find that the effect of NO$_2$ adsorption is unstable against photon exposure. This result suggests that photo-induced doping could be an additional way with which to tune the electron-doped surface states on Bi$_2$Te$_3$(111). We present the photon-induced evolution of the surface band structure data in
Fig.~\ref{fig:fig3}. To reduce the uncertainties due to photon-related effects, each spectrum was collected within 1 minute after opening the photon shutter. Fig.~\ref{fig:fig3}(a) presents the surface band dispersion after the sample was first exposed to 0.3L of NO$_2$. The resulting Dirac point is at about 330meV below
E$_F$. With photon exposure the chemical potential gradually shifts upwards. After a dosage of 1$\times$$10^{15}$ photons 30eV, the Dirac point is shifted to approximately 250meV below E$_F$. Successive exposures continue to shift the chemical potential. After a total dosage of approximately 8$\times$$10^{15}$ photons per mm$^2$, the chemical potential is found to be stabilized as the Dirac point reaches 130meV, the energy location prior to the NO$_2$ exposure. The photon-induced effective doping is observed for a wide range of UV energies (28-55 eV) and is found to be robustly reproducible.

For comparison with other topological insulators that may have a different chemically-active surface with different relaxation details, we present the effect of photo-induced doping on the Bi$_2$Se$_3$ (111) surface doped with NO$_2$ molecules (Fig.~\ref{fig:fig3}(g)-(l)). Previous systematic study \cite{qian} has shown that exposure to NO$_2$ dopants hole-dopes the Bi$_2$Se$_3$ surface states, moving the chemical potential to the Dirac point. While this behavior is opposite to that observed for our Bi$_2$Te$_3$ (telluride system), the additional effect of photodoping is similar for both systems. In Figure~\ref{fig:fig3}(g) we present the surface band dispersion data taken after a fixed dosage of 0.1L of NO$_2$. The resulting Dirac point is at a binding energy of approximately 150meV. Similar to Bi$_2$Te$_3$, photo-induced doping gradually moves the chemical potential downwards. A dosage of 2$\times$$10^{13}$ photons at 30eV shifts the Dirac point by 50meV, while a total dosage of 6$\times$$10^{14}$ photons shifts the Dirac point to the Fermi level (Fig.~\ref{fig:fig3}(l)). While the effect of photo-induced doping in Bi$_2$Te$_3$ could be attributed to photo-stimulated desorption, the
detailed mechanism in Bi$_2$Se$_3$ is still under investigation by us. Nevertheless, we have demonstrated that photon-induced hole doping can be a controllable mechanism to manipulate the topological spin-polarized surface bands, systematically driving the chemical potential to the Dirac point via photon flux.

In summary, using high resolution ARPES, we have demonstrated for the first time
a suite of methods to controllably manipulate the topological edge states of the
topological insulator Bi$_2$Te$_3$ on its (111)-surface without breaking time reversal symmetry. Our systematic data (Fig.1-3) directly show that both potassium (K) and NO$_2$ electron-dopes the topologically ordered surface over a wide doping-range. Additionally, we showed that the effect of the dosage can be reversed by controlling exposure to photon-flux. We further showed that photodoping has a similar effect on Bi$_2$Se$_3$, which can be used to place the chemical potential to the much desired Dirac point. Our work opens up many new possibilities for future utility of topological insulators in spintronic and quantum computing research. More importantly, our methods of topological surface state manipulation are highly suitable for spectroscopic studies than the traditional electrical gating based techniques thus open up new spectroscopic investigation opportunities of topological quantum phenomena.

NOTE ADDED: Some of the methods and their combinations demonstrated here can be successfully applied to other topological insulators such as the bulk-insulating-Bi$_{1-x}$Sb$_x$, Sb$_2$Te$_3$ and Bi$_2$Se$_3$ by carefully considering their unique material-specific surface chemistry and the nature of the surface relaxation which will be shown elsewhere \cite{qian}.

\begin{figure}
\includegraphics[width=.80\textwidth]{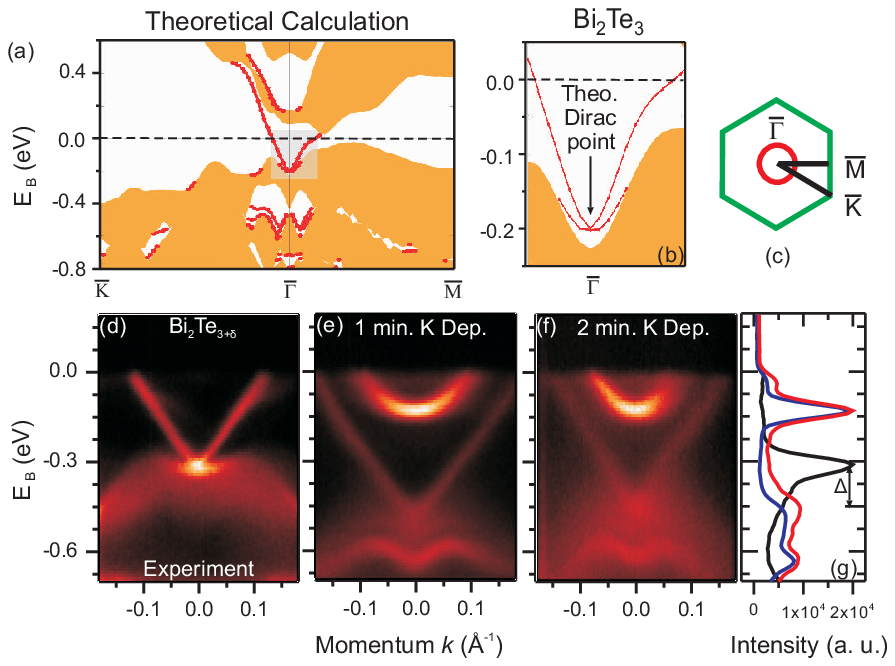}
\caption{(Enlarged version of Fig-1) \label{fig:fig1} \textbf{Electron doping via potassium (K) deposition on Bi$_2$Te$_3$[111] topological surface :} (a)-(b) (111) Surface band (red lines) calculation with spin-orbit coupling shows a pair of non-degenerate Dirac bands
crossing the E$_F$. The bulk band projection is denoted by the shaded
region in orange. The resulting (c) Fermi surface is a single non-degenerate ring
centered at the zone center. (d) ARPES data on Bi$_{2}$Te$_{3+\delta}$ shows a slight rise of the chemical potential relative to the stoichiometric compound Bi$_{2}$Te$_{3}$ \cite{Hsieharxiv}. The surface band structure after depositing potassium (K) for (e) one minute and (f) two minutes show additional electron-doping of the surface
states. The (g) energy distribution curves at $\bar{\Gamma}$ show
that relative to the pristine sample (black), the chemical potential
is shifted by 140meV after one minute of (red) of K deposition. Only a small additional shift of the E$_F$ position is observed upon an additional minute of K deposition (blue).}
\end{figure}

\begin{figure*}

\includegraphics[width=0.8\textwidth]{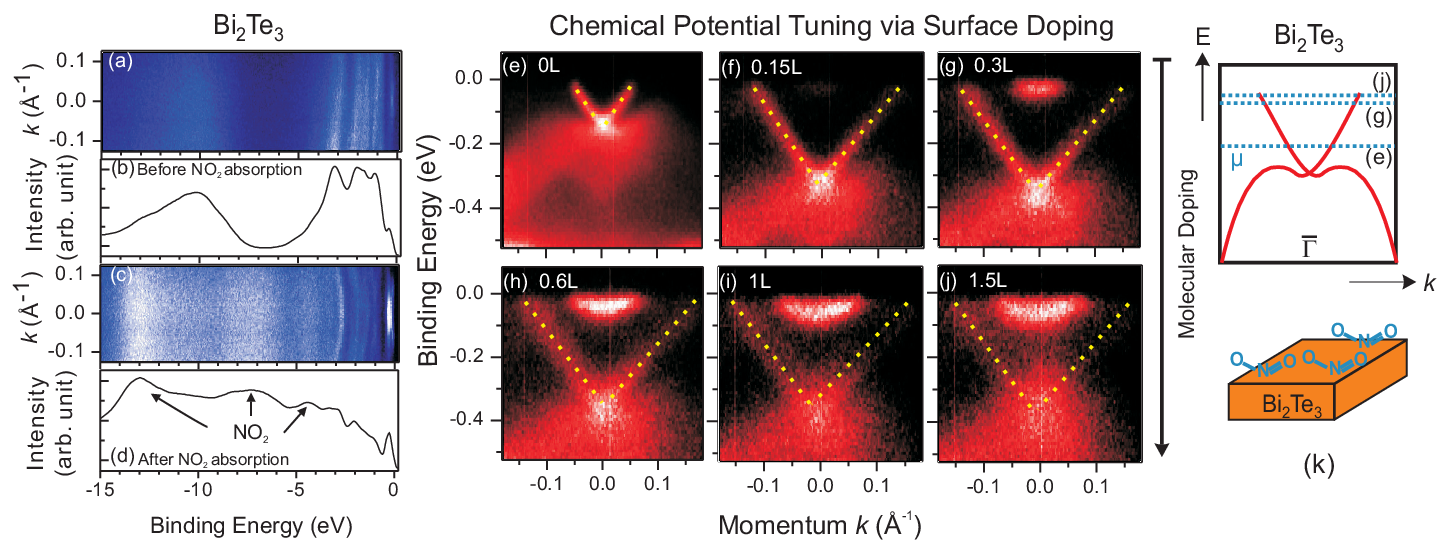}
\caption{(Enlarged version of Fig-2)\label{fig:fig2} \textbf{Manipulation of Bi$_2$Te$_3$[111] topological surface via molecular adsorption:} (a) Angle-resolved valence band spectra (a)
prior to and (c) post exposure to 1.5 Langmuirs of NO$_2$. Panels (b)
and (d) show the angle integrated spectra for (a) and (c),
respectively. Prior to the NO$_2$ adsorption, the valence band features are sharp
and dispersive with momentum. After exposure to NO$_2$ gas, three non-dispersive
features appear at binding energies of 4.4eV, 7.5eV and 13eV. High
resolution surface band dispersion data near the $\bar{\Gamma}$ point
suggest (e) two non-degenerate surface bands intersecting at 130meV
below E$_F$, forming a Dirac point. (f)-(j), with NO$_2$ adsorption,
the chemical potential is systematically raised, thereby introducing additional
electrons into the Bi$_2$Te$_3$ surface bands. The Dirac point moves
away from E$_F$ with increasing amounts of NO$_2$ exposure. At 1
Langmuir dosage value, the lowest surface conduction band gains electron occupation, and the Dirac point reaches 350meV below E$_F$.}
\end{figure*}

\begin{figure*}
\includegraphics[width=1.05\textwidth]{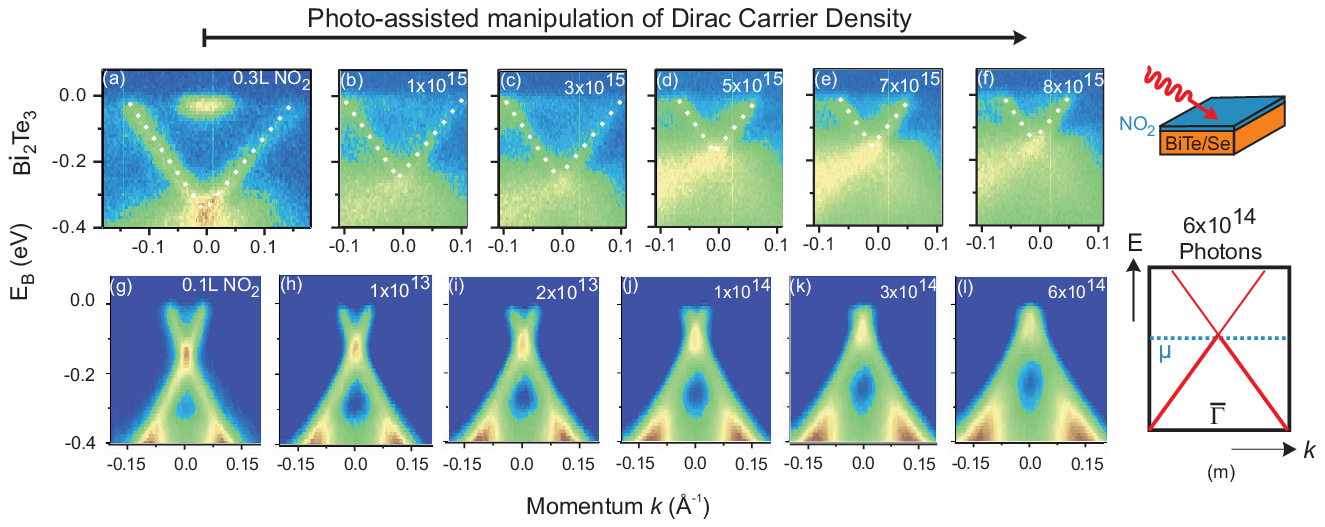}
\caption{(Enlarged version of Fig-3)\label{fig:fig3} \textbf{Photo-induced manipulation of the topological surface
states on Bi$_2$Te$_3$:} (a) Upon exposures of 0.3L of NO$_2$ on Bi$_2$Te$_3$, the
Dirac point is located at approximately 330meV below E$_F$. (b)-(f) Photon
exposure reverses the electron transfer to the Bi$_2$Te$_3$ surface,
moving the surface chemical potential in a well-controlled manner
shifting towards the Dirac point. With a dosage total of (b) 1$\times$$10^{15}$ photons, the Dirac point is moved to 250meV below E$_F$. The Dirac point
eventually reaches the energy-location prior to the NO$_2$ exposure with a dosage
of 8$\times$$10^{15}$ photons, which is at about 130meV below E$_F$. For comparison, the
photo-induced doping of Bi$_2$Se$_3$ surface states exposed with (g) 0.1L of
NO$_2$ gas is also presented in (h)-(l). With a dosage level of
(i)2$\times$$10^{13}$ photons, the Dirac point is shifted by 50meV to
to 100meV below E$_F$. The chemical potential is eventually stabilized at the Dirac point by the application of $6\times$$10^{14}$ photons per mm$^2$.}
\end{figure*}

\end{document}